\documentclass[twocolumn,american,portuges,spanish,brazil,10pt,twocolumn,brazil,aps,superscriptaddress, subscriptfootnote,nofootinbib]{revtex4}
\usepackage[T1]{fontenc}
\usepackage[latin9]{inputenc}
\usepackage{color}
\usepackage{babel}
\addto\shorthandsspanish{\spanishdeactivate{~<>}}

\usepackage{float}
\usepackage{amsmath}
\usepackage{amssymb}
\usepackage{graphicx}
\usepackage[unicode=true,pdfusetitle,
 bookmarks=true,bookmarksnumbered=false,bookmarksopen=false,
 breaklinks=true,pdfborder={0 0 0},backref=false,colorlinks=true]
 {hyperref}
\usepackage{breakurl}

\makeatletter
\@ifundefined{textcolor}{}
{%
 \definecolor{BLACK}{gray}{0}
 \definecolor{WHITE}{gray}{1}
 \definecolor{RED}{rgb}{1,0,0}
 \definecolor{GREEN}{rgb}{0,1,0}
 \definecolor{BLUE}{rgb}{0,0,1}
 \definecolor{CYAN}{cmyk}{1,0,0,0}
 \definecolor{MAGENTA}{cmyk}{0,1,0,0}
 \definecolor{YELLOW}{cmyk}{0,0,1,0}
}

\hypersetup{colorlinks=true,citecolor=blue,linkcolor=cyan,urlcolor=blue,filecolor= green}

\makeatother

\begin{document}

\title{A representa\c{c}\~ao de Kraus para a din\^amica de sistemas qu\^anticos
abertos \\
\foreignlanguage{american}{ {\small{}{(The Kraus representation for
the dynamics of open quantum systems)}}}}

\author{Jonas Maziero}

\email{jonas.maziero@ufsm.br}

\address{Departamento de F\'isica, Centro de Ci\^encias Naturais e Exatas, Universidade Federal de Santa Maria, Avenida Roraima 1000, 97105-900, Santa Maria, RS, Brazil}

\selectlanguage{spanish}%

\address{Instituto de F\'isica, Facultad de Ingenier\'ia, Universidad de la Rep\'ublica, J. Herrera y Reissig 565, 11300, Montevideo, Uruguay}
\selectlanguage{brazil}%
\begin{abstract}
A necessidade e utilidade de se considerar a intera\c{c}\~ao  com
o ambiente quando da descri\c{c}\~ao da evolu\c{c}\~ao temporal
de sistemas qu\^anticos vem sendo reconhecida nos mais variados ramos
da f\'isica e de outras ci\^encias. A representa\c{c}\~ao de Kraus \'e
uma forma geral e sucinta para descrever a din\^amica de sistemas
qu\^anticos abertos em muitas situa\c{c}\~oes f\'isicas relevantes.
Neste artigo, nos abstendo da generalidade do formalismo de opera\c{c}\~oes qu\^anticas
e evitando assim as complica\c{c}\~oes associadas, mostramos de forma
simples como obter tal representa\c{c}\~ao usando basicamente a evolu\c{c}\~ao unit\'aria
do sistema fechado (sistema mais ambiente) e a fun\c{c}\~ao tra\c{c}o
parcial. O exemplo de um \'atomo de dois n\'iveis interagindo com
o v\'acuo do campo eletromagn\'etico \'e considerado para ilustrar
a aplica\c{c}\~ao desse formalismo, que por fim \'e utilizado para
estudar a evolu\c{c}\~ao temporal da coer\^encia qu\^antica do \'atomo.
\\
 \textbf{Palavras chave:} mec\^anica qu\^antica, sistemas abertos,
representa\c{c}\~ao de Kraus \\

\vspace{0.01mm}

\selectlanguage{american}%
The necessity and utility of considering the interaction of a quantum
system with its environment when describing its time evolution have
been recognized in several branches of physics and of other sciences.
The Kraus' representation is a general and succinct approach to describe
such open system dynamics in a wide range of relevant physical scenarios.
In this article, by abdicating from the generality of the formalism
of quantum operations and with this avoiding its associated complications,
we show in a simple manner how we can obtain the Kraus' representation
using basically the closed system (system plus environment) unitary
dynamics and the partial trace function. The example of a two-level
atom interacting with the vacuum of the electromagnetic field is regarded
for the sake of instantiating this formalism, which is then applied
to study the time evolution of the atom's quantum coherence.
\end{abstract}
\maketitle

\section{Introdu\c{c}\~ao}

A mec\^anica qu\^antica (MQ) \'e um dos grandes triunfos obtidos
na constru\c{c}\~ao do conhecimento humano e \'e a base sobre a
qual se desenvolveu boa parte das tecnologias atuais \cite{Kakalios}.
Hoje em dia, informa\c{c}\~ao possui um papel fundamental em nossas
vidas e por isso dizemos que estamos na era da informa\c{c}\~ao.
Ainda nesse s\'eculo entraremos em outra etapa da nossa hist\'oria,
a \emph{era da informa\c{c}\~ao qu\^antica }\cite{IOP,Laflamme}.
Testemunharemos a realiza\c{c}\~ao de experimentos controlando propriedades
f\'isicas tais como coer\^encia e correla\c{c}\~oes qu\^anticas
de forma que jamais foi cogitada pelos precursores da MQ (veja e.g.
as Refs. \cite{Rodrigues}, p. 70, e \cite{Waerden}), nem pela maioria
de n\'os hoje em dia. Mesclando essa habilidade com a engenhosidade
instigada pela corrida tecnol\'ogica, lograremos construir m\'aquinas
e dispositivos que ter\~ao um impacto inimagin\'avel na maneira
como vivemos e interagimos.

Embora haja muito ainda por ser feito, j\'a conseguimos obter avan\c{c}os
importantes em v\'arias sub-\'areas do que chamamos atualmente de\emph{
ci\^encia da informa\c{c}\~ao qu\^antica }(CIQ), mas que na verdade
engloba muitos ramos da ci\^encia. Investimentos substanciais est\~ao
sendo feitos, por exemplo, para construir computadores e simuladores
qu\^anticos que poder\~ao resolver alguns problemas matem\'aticos
e simular o comportamento de sistemas complexos de maneira muito mais
eficiente que os computadores convencionais, cl\'assicos, o fazem
\cite{OBrien,Nori}. Este \'e s\'o um de diversos exemplos de iniciativas
promissoras em CIQ, que v\~ao desde a possibilidade de comunica\c{c}\~ao
absolutamente segura \cite{Ekert} e de medidas mais precisas em metrologia
\cite{LloydM,DavidovichM}, passando pela termodin\^amica qu\^antica
de sistemas ``pequenos'' \cite{Jarzynski} e pelo entendimento de
e inspira\c{c}\~ao em sistemas biol\'ogicos \cite{McFadden} e indo
at\'e a ramos t\~ao abrangentes quanto a intelig\^encia artificial
\cite{PetruccioneAI,PetruccioneNN,Svore,Andraca}.

Tendo essa perspectiva em vistas, devemos reconhecer a import\^ancia
de prepararmos e educarmos nossas crian\c{c}as e jovens para que
esse ``novo mundo'' seja mais natural (menos estranho) para eles do
que \'e para n\'os. Por isso, iniciativas objetivando facilitar
o entendimento da MQ s\~ao necess\'arias \cite{Caltech,Wiesner,Greca,Wieman,Arndt,QiCloud}.
A motiva\c{c}\~ao para este artigo vem da observa\c{c}\~ao de que
na maioria dos livros texto sobre MQ, tipicamente se assume, implicitamente,
que o sistema f\'isico de interesse \'e fechado, ou seja, que n\~ao
interage como o seu ambiente e que sua din\^amica \'e unit\'aria.
Aqui esperamos contribuir para o ensino de MQ apresentando de maneira
simples a representa\c{c}\~ao de Kraus para a din\^amica qu\^antica
de \emph{sistemas abertos}, o que \'e o caso para a maioria dos sistemas
f\'isicos reais.

Uma maneira interessante para iniciar o aprendizado mais aprofundado
de MQ \'e, depois de ter obtido uma boa base de \'algebra linear
e de ter feito leituras sobre a fenomenologia da MQ, estudar seus
postulados fundamentais \cite{Zurek,Nielsen=000026Chuang,Preskill,Wilde,MazieroRBEF}.
Um apresenta\c{c}\~ao did\'atica dos postulados dos estados e das
medidas pode ser encontrada na Ref. \cite{MazieroRBEF}. Aqui focaremos
no postulado da din\^amica. Antes de apresent\'a-lo, no entanto,
lembremos que o estado de um sistema f\'isico \'e descrito, de forma
mais geral, por um operador densidade, que \'e um operador linear,
positivo semi-definido e que possui tra\c{c}o igual a um\footnote{Um operador $\hat{O}$ \'e linear se $\hat{O}(\sum_{j}c_{j}|\psi_{j}\rangle)=\sum_{j}c_{j}\hat{O}(|\psi_{j}\rangle)$,
onde $c_{j}\in\mathbb{C}$ e $\{|\psi_{j}\rangle\}$ \'e um conjunto
qualquer de vetores do espa\c{c}o de Hilbert $\mathcal{H}$ \cite{MazieroRBEF}.
Esse operador \'e positivo semi-definido (nota\c{c}\~ao $\hat{O}\ge\hat{0}$,
onde $\hat{0}$ \'e o operador nulo) se $\langle\psi|\hat{O}|\psi\rangle\ge0$
para todo vetor $|\psi\rangle\in\mathcal{H}$. O tra\c{c}o de $\hat{O}$
\'e definido como: $\mathrm{Tr}(\hat{O})=\sum_{j=1}^{d}\langle\xi_{j}|\hat{O}|\xi_{j}\rangle$,
com $d$ sendo a dimens\~ao e $\{|\xi_{j}\rangle\}$ uma base de
$\mathcal{H}$. }. 

O \textit{postulado da din\^amica} diz que, se $\hat{H}$ \'e o
operador hamiltoniano (energia) de um sistema isolado no instante
de tempo $t$, seu estado evolui no tempo atrav\'es de uma transforma\c{c}\~ao unit\'aria,
ou seja,
\begin{equation}
\hat{\rho}_{t}=\hat{U}\hat{\rho}\hat{U}^{\dagger},\label{eq:rhoTU}
\end{equation}
em que $\hat{\rho}$ \'e o estado inicial (em $t=0$) do sistema
e o operador de evolu\c{c}\~ao temporal \'e obtido a partir da equa\c{c}\~ao de Schr\"odinger:
\begin{equation}
i\hbar\partial_{t}\hat{U}=\hat{H}\hat{U},\label{eq:eqS}
\end{equation}
em que $\hbar$ \'e a constante de Planck e, neste artigo, usaremos
a nota\c{c}\~ao $\partial_{t}\equiv\frac{\partial}{\partial t}.$
N\~ao \'e dif\'icil verificar que as Eqs. (\ref{eq:rhoTU}) e (\ref{eq:eqS})
s\~ao equivalentes \`a equa\c{c}\~ao de von Neumann\footnote{No caso de estados iniciais puros, i.e., $\hat{\rho}=|\psi\rangle\langle\psi|$
(quando n\~ao existe incerteza cl\'assica sobre a prepara\c{c}\~ao
do sistema \cite{MazieroRBEF}), temos $|\psi_{t}\rangle=\hat{U}|\psi\rangle$,
que, junto com a Eq. (\ref{eq:rhoTU}), \'e equivalente \`a equa\c{c}\~ao
de Schr\"odinger para o estado: $i\hbar\partial_{t}|\psi_{t}\rangle=\hat{H}|\psi_{t}\rangle$.}:
\begin{equation}
i\hbar\partial_{t}\hat{\rho}_{t}=[\hat{H},\hat{\rho}_{t}].\label{eq:eqvN}
\end{equation}
Vale mencionar que, por suas estruturas alg\'ebricas, as Eqs. (\ref{eq:rhoTU})
e (\ref{eq:eqvN}) s\~ao chamadas, respectivamente, de vers\~oes
discreta\footnote{Pois, uma vez que $\hat{U}$ \'e conhecido, essa equa\c{c}\~ao \'e
uma mapa discreto $\hat{\rho}\mapsto\hat{\rho}_{t}$.} e cont\'inua\footnote{Pois essa equa\c{c}\~ao \'e uma equa\c{c}\~ao diferencial que envolve
varia\c{c}\~oes de $\hat{\rho}$ em intervalos de tempo infinitesimais
em torno do instante de tempo $t$.} da din\^amica qu\^antica de sistemas fechados. Neste artigo trabalharemos
com a generaliza\c{c}\~ao para sistemas abertos da vers\~ao discreta,
que \'e devida a Kraus \cite{Kraus_AP,Kraus_B} e \'e bastante \'util
para descrever a evolu\c{c}\~ao temporal de v\'arios sistemas f\'isicos
reais. Para mais informa\c{c}\~oes sobre generaliza\c{c}\~oes da
vers\~ao cont\'inua, tais como os formalismos de Redfield, das equa\c{c}\~oes
mestras e das trajet\'orias qu\^anticas, veja \cite{Fanchini_eqM,Mizrahi,DavidovichQJ,Celeri_RedField}
e suas refer\^encias.

Quando lidamos com dois (ou mais) graus de liberdade, chamados e.g.
de $S$ e $A$, ou seja, quando consideramos \emph{sistemas compostos},
complementamos o postulado dos estados \cite{MazieroRBEF} assumindo
que o estado do sistema como um todo \'e descrito por um operador
densidade definido no espa\c{c}o de Hilbert que \'e obtido tomando-se
o produto tensorial dos espa\c{c}os individuais, i.e., $\mathcal{H}_{S}\otimes\mathcal{H}_{A}=:\mathcal{H}_{SA}$
\cite{Zurek}. Neste contexto, frequentemente conhecemos o estado
global mas queremos calcular probabilidades relacionadas a somente
um dos sub-sistemas. Para isso \'e conveniente usar a fun\c{c}\~ao
linear\emph{ tra\c{c}o parcial}, cuja defini\c{c}\~ao \'e \cite{Nielsen=000026Chuang}:
\begin{eqnarray}
\mathrm{Tr}_{A}\left(|\psi\rangle\langle\phi|\otimes|\xi\rangle\langle\zeta|\right) & := & |\psi\rangle\langle\phi|\otimes\mathrm{Tr}_{A}\left(|\xi\rangle\langle\zeta|\right)\nonumber \\
 & = & \langle\zeta|\xi\rangle|\psi\rangle\langle\phi|,
\end{eqnarray}
em que $|\psi\rangle,|\phi\rangle\in\mathcal{H}_{S}$ e $|\xi\rangle,|\zeta\rangle\in\mathcal{H}_{A}$
e tomamos o tra\c{c}o sobre $\mathcal{H}_{A}$. Assim, por exemplo,
se o estado do sistema bipartido $SA$ \'e $\hat{\rho}$, o estado
do sistema $S$ pode ser obtido usando-se o tra\c{c}o parcial:
\[
\hat{\rho}^{S}=\mathrm{Tr}_{A}\left(\hat{\rho}\right).
\]

A continua\c{c}\~ao deste artigo est\'a organizada da seguinte forma.
Come\c{c}aremos, na Sec. \ref{demKOSR}, obtendo a representa\c{c}\~ao
de Kraus para a din\^amica de um sistema qu\^antico aberto. Na sequ\^encia,
na Sec. \ref{sec_atom}, forneceremos um exemplo de aplica\c{c}\~ao
dessa representa\c{c}\~ao na descri\c{c}\~ao da evolu\c{c}\~ao
temporal de um \'atomo de dois n\'iveis, e de sua coer\^encia qu\^antica,
quando esse interage com o campo eletromagn\'etico inicialmente no
estado de v\'acuo. Deixamos para o \hyperref[propriedades]{Ap\^endice}
as provas das v\'arias propriedades que s\~ao exigidas (ou esperadas)
para os operadores e a din\^amica de Kraus.

\section{\textmd{\textup{\normalsize{}A }}representa\c{c}\~ao de Kraus a
partir da din\^amica unit\'aria}

\label{demKOSR}

\begin{figure}[b]
\begin{centering}
\includegraphics[scale=0.77]{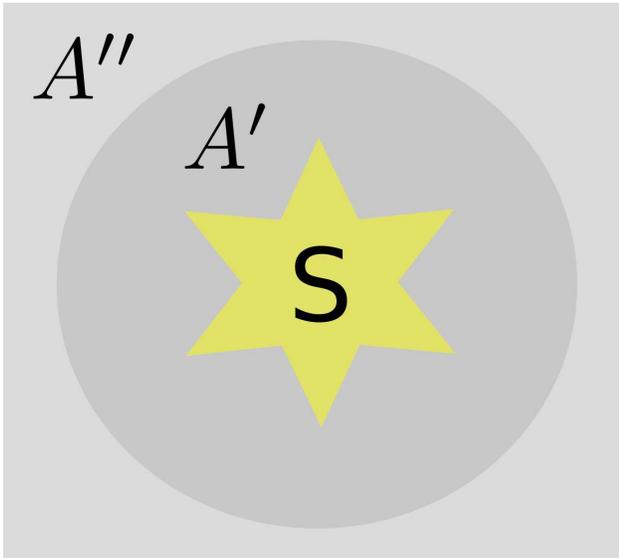}
\par\end{centering}

\caption{O sistema $S$, preparado no estado $\hat{\rho}^{S}$, pode ser influenciado
efetivamente pelo ambiente $A'$. Se a decomposi\c{c}\~ao espectral
do estado inicial de $A'$ \'e $\hat{\rho}_{A'}=\sum_{j}a_{j}'|a_{j}'\rangle\langle a_{j}'|$,
podemos utilizar a decomposi\c{c}\~ao de Schmidt \cite{Nielsen=000026Chuang}
e outro ambiente $A''$ para purificar esse estado: $\hat{\rho}_{A'}=\mathrm{Tr}_{A''}\left(|E_{0}\rangle\langle E_{0}|\right)$,
com $|E_{0}\rangle:=\sum_{j}\sqrt{a_{j}'}|a_{j}'\rangle\otimes|a_{j}''\rangle\in\mathcal{H}_{A'}\otimes\mathcal{H}_{A''}\equiv\mathcal{H}_{A'A''}=:\mathcal{H}_{A}$
e $\{|a_{j}''\rangle\}$ \'e uma base ortonormal de $\mathcal{H}_{A''}$.}

\label{fig_syst_env}
\end{figure}

Nesta se\c{c}\~ao obteremos, de maneira simples e sem ambiguidades,
a representa\c{c}\~ao de Kraus para o estado evolu\'ido de um sistema
qu\^antico discreto. Tentamos simplificar ao m\'aximo a nota\c{c}\~ao
para que o material seja acess\'ivel a todos os que possuem um conhecimento
b\'asico do formalismo matem\'atico da MQ de sistemas fechados.
Para evitar as complica\c{c}\~oes relacionadas \`a quest\~ao da poss\'ivel n\~ao
positividade completa da din\^amica qu\^antica gerada a partir de
estados iniciais correlacionados \cite{Pechukas,Alicki,Shaji,Rosario,Shabani,Brodutch,Buscemi},
vamos considerar que sistema e ambiente est\~ao inicialmente em um
estado produto:
\begin{equation}
\hat{\rho}=\hat{\rho}^{S}\otimes|E_{0}\rangle\langle E_{0}|.
\end{equation}
O ambiente propriamente dito, ou seja, a parte do universo que pode
interagir efetivamente com o sistema $S$ em uma escala de tempo relevante,
 poderia estar em um estado misto qualquer. No entanto, para facilitar
as demonstra\c{c}\~oes, vamos utilizar uma purifica\c{c}\~ao deste
estado \cite{Nielsen=000026Chuang,Preskill,Wilde}; veja a Fig. \ref{fig_syst_env}.

O primeiro passo para a obten\c{c}\~ao da representa\c{c}\~ao de Kraus \'e
considerar o sistema $S$ mais seu ambiente $A$ como sendo um sistema
fechado. Assim, o estado  desse sistema composto ser\'a evolu\'ido
no tempo atrav\'es de uma transforma\c{c}\~ao unit\'aria, como
mostrado na Eq. (\ref{eq:rhoTU}), com $\hat{U}$ obtido da equa\c{c}\~ao de Schr\"odinger,
Eq. (\ref{eq:eqS}), com
\begin{equation}
\hat{H}=\hat{H}_{S}\otimes\mathbb{I}_{A}+\mathbb{I}_{S}\otimes\hat{H}_{A}+g\hat{H}_{int},
\end{equation}
onde $\mathbb{I}_{S(A)}$ \'e o operador identidade no espa\c{c}o
vetorial correspondente, $\hat{H}_{S(A)}$ \'e o chamado Hamiltoniano
livre do sistema (ambiente) e $g$ \'e um par\^ametro real que determina
o ``qu\~ao forte'' \'e a intera\c{c}\~ao entre o sistema e seu
ambiente, cuja forma \'e descrita pelo Hamiltoniano de intera\c{c}\~ao
$\hat{H}_{int}$. Se $g=0$ sistema e ambiente evoluem de forma independente,
e continuam descorrelacionados. Por outro lado, se $g\ne0$ correla\c{c}\~oes s\~ao
geradas entre sistema e ambiente \cite{Maziero12,Walborn12}, o que
pode influenciar as caracter\'istcas qu\^anticas de $S$ causando,
por exemplo, perda de coer\^encia qu\^antica em um processo conhecido
como decoer\^encia \cite{Schlosshauer,Zurek_Dar}.

O leitor certamente j\'a verificou que operadores unit\'arios preservam
o produto escalar entre vetores do espa\c{c}o no qual atuam e que
o efeito dessas transforma\c{c}\~oes \'e  simplesmente o de realizar
uma mudan\c{c}a de base, ou de representa\c{c}\~ao. Ent\~ao, se
consideramos duas bases quaisquer de vetores ortonormais do espa\c{c}o
de Hilbert para o sistema composto,
\begin{equation}
\{|\Phi_{j}\rangle\},\{|\Psi_{j}\rangle\}\in\mathcal{H}_{S}\otimes\mathcal{H}_{A},
\end{equation}
com $j=0,\cdots,d_{S}d_{A}-1$, onde $d_{S(A)}=\mathrm{dim}(\mathcal{H}_{S(A)})$
\'e a dimens\~ao do espa\c{c}o de estados para o sistema (ambiente),
podemos escrever:
\begin{equation}
\hat{U}=\sum_{j}|\Phi_{j}\rangle\langle\Psi_{j}|.\label{eq:Uent}
\end{equation}

As bases que aparecem na \'ultima equa\c{c}\~ao definem a a\c{c}\~ao
do operador unit\'ario e s\~ao completamente gerais, podendo n\~ao
ser um simples produto tensorial de bases de $\mathcal{H}_{S}$ e
de $\mathcal{H}_{A}$, fato que depender\'a da forma do Hamiltoniano
de intera\c{c}\~ao. Ser\'a \'util, para alcan\c{c}armos nosso objetivo
aqui, escrever os vetores dessas duas bases gerais em termos de uma
base produto:
\begin{equation}
\{|S_{k}\rangle\otimes|E_{l}\rangle=:|S_{k}E_{l}\rangle\},
\end{equation}
onde $k=0,\cdots,d_{S}-1$ and $l=0,\cdots,d_{A}-1$. A rela\c{c}\~ao
de completeza em $\mathcal{H}_{SA}$,
\begin{equation}
\mathbb{I}=\sum_{kl}|S_{k}E_{l}\rangle\langle S_{k}E_{l}|,
\end{equation}
\'e ent\~ao utilizada para escrever 
\begin{eqnarray}
|\Phi_{j}\rangle & = & \mathbb{I}|\Phi_{j}\rangle\\
 & = & \sum_{kl}|S_{k}E_{l}\rangle\langle S_{k}E_{l}||\Phi_{j}\rangle\\
 & := & \sum_{kl}f_{kl}^{(j)}|S_{k}E_{l}\rangle,
\end{eqnarray}
onde definimos os coeficientes complexos
\begin{equation}
f_{kl}^{(j)}:=\langle S_{k}E_{l}|\Phi_{j}\rangle.
\end{equation}

Analogamente
\begin{equation}
|\Psi_{j}\rangle=\sum_{kl}g_{kl}^{(j)}|S_{k}E_{l}\rangle,
\end{equation}
com os coeficientes
\begin{equation}
g_{kl}^{(j)}:=\langle S_{k}E_{l}|\Psi_{j}\rangle.
\end{equation}

Assim o operador unit\'ario toma a forma: 
\begin{eqnarray}
\hat{U} & = & \sum_{j}\sum_{kl}f_{kl}^{(j)}|S_{k}E_{l}\rangle\sum_{mn}g_{mn}^{(j)*}\langle S_{m}E_{n}|\\
 & = & \sum_{jklmn}f_{kl}^{(j)}g_{mn}^{(j)*}|S_{k}E_{l}\rangle\langle S_{m}E_{n}|,
\end{eqnarray}
em que usamos o fato de que $(c|\xi\rangle)^{\dagger}=c^{*}\langle\xi|$
para qualquer $c\in\mathbb{C}$, com $c^{*}$ sendo o complexo conjugado
de $c$.

Usando o operador unit\'ario escrito em termos da base produto, aplicado
como mostrado na Eq. (\ref{eq:rhoTU}), e a fun\c{c}\~ao tra\c{c}o
parcial, o estado evolu\'ido do sistema $S$,
\begin{equation}
\hat{\rho}_{t}^{S}=\mathrm{Tr}_{A}(\hat{\rho}_{t})=\mathrm{Tr}_{A}(\hat{U}\hat{\rho}\hat{U}^{\dagger}),
\end{equation}
\'e obtido como segue:

\begin{widetext}

\begin{eqnarray}
\hat{\rho}_{t}^{S} & = & \mathrm{Tr}_{A}\sum_{jklmn}f_{kl}^{(j)}g_{mn}^{(j)*}|S_{k}E_{l}\rangle\langle S_{m}E_{n}|\left(\hat{\rho}^{S}\otimes|E_{0}\rangle\langle E_{0}|\right)\sum_{opqrs}f_{pq}^{(o)*}g_{rs}^{(o)}|S_{r}E_{s}\rangle\langle S_{p}E_{q}|\nonumber \\
 & = & \mathrm{Tr}_{A}\sum_{jklmnopqrs}f_{kl}^{(j)}g_{mn}^{(j)*}f_{pq}^{(o)*}g_{rs}^{(o)}|S_{k}E_{l}\rangle\langle S_{m}E_{n}|\left(\hat{\rho}^{S}\otimes|E_{0}\rangle\langle E_{0}|\right)|S_{r}E_{s}\rangle\langle S_{p}E_{q}|\label{eq:pt1}\\
 & = & \mathrm{Tr}_{A}\sum_{jklmnopqrs}f_{kl}^{(j)}g_{mn}^{(j)*}f_{pq}^{(o)*}g_{rs}^{(o)}|S_{k}\rangle\langle S_{m}|\hat{\rho}^{S}|S_{r}\rangle\langle S_{p}|\otimes|E_{l}\rangle\underset{=\delta_{n0}}{\underbrace{\langle E_{n}|E_{0}\rangle}}\underset{=\delta_{0s}}{\underbrace{\langle E_{0}|E_{s}\rangle}}\langle E_{q}|\label{eq:pt2}\\
 & = & \sum_{jklmopqr}f_{kl}^{(j)}g_{m0}^{(j)*}f_{pq}^{(o)*}g_{r0}^{(o)}|S_{k}\rangle\langle S_{m}|\hat{\rho}^{S}|S_{r}\rangle\langle S_{p}|\underset{=\langle E_{q}|E_{l}\rangle=\delta_{ql}}{\underbrace{\mathrm{Tr}_{A}(|E_{l}\rangle\langle E_{q}|)}}\label{eq:pt3}\\
 & = & \sum_{jklmopr}f_{kl}^{(j)}g_{m0}^{(j)*}f_{pl}^{(o)*}g_{r0}^{(o)}|S_{k}\rangle\langle S_{m}|\hat{\rho}^{S}|S_{r}\rangle\langle S_{p}|\label{eq:pt4}\\
 & = & \sum_{jklmopr}\langle S_{k}E_{l}|\Phi_{j}\rangle\langle\Psi_{j}|S_{m}E_{0}\rangle\langle\Phi_{o}|S_{p}E_{l}\rangle\langle S_{r}E_{0}|\Psi_{o}\rangle|S_{k}\rangle\langle S_{m}|\hat{\rho}^{S}|S_{r}\rangle\langle S_{p}|\label{eq:pt5}\\
 & = & \sum_{jklmopr}\langle S_{k}E_{l}|\Phi_{j}\rangle\langle\Psi_{j}|S_{m}E_{0}\rangle|S_{k}\rangle\langle S_{m}|\hat{\rho}^{S}|S_{r}\rangle\langle S_{p}|\langle S_{r}E_{0}|\Psi_{o}\rangle\langle\Phi_{o}|S_{p}E_{l}\rangle\label{eq:pt6}\\
 & = & \sum_{klmpr}\langle S_{k}E_{l}|\left({\textstyle \sum_{j}}|\Phi_{j}\rangle\langle\Psi_{j}|\right)|S_{m}E_{0}\rangle|S_{k}\rangle\langle S_{m}|\hat{\rho}^{S}|S_{r}\rangle\langle S_{p}|\langle S_{r}E_{0}|\left({\textstyle \sum_{o}}|\Psi_{o}\rangle\langle\Phi_{o}|\right)|S_{p}E_{l}\rangle\label{eq:pt7}\\
 & = & \sum_{klmpr}\langle S_{k}E_{l}|\hat{U}|S_{m}E_{0}\rangle|S_{k}\rangle\langle S_{m}|\hat{\rho}^{S}|S_{r}\rangle\langle S_{p}|\langle S_{r}E_{0}|\hat{U}^{\dagger}|S_{p}E_{l}\rangle.\label{eq:pt8}
\end{eqnarray}

\end{widetext}

Acima $\delta_{jk}$ \'e a fun\c{c}\~ao delta de Kronecker. Na passagem
da Eq. (\ref{eq:pt1}) para a Eq. (\ref{eq:pt2}) lembramos que $|S_{k}E_{l}\rangle=|S_{k}\rangle\otimes|E_{l}\rangle$
e usamos o fato de que
\begin{equation}
(\hat{A}\otimes\hat{B})(\hat{C}\otimes\hat{D})=\hat{A}\hat{C}\otimes\hat{B}\hat{D}
\end{equation}
para quaisquer quatro matrizes complexas $\hat{A},\hat{B},\hat{C},\hat{D}$.
Para obter (\ref{eq:pt3}) de (\ref{eq:pt2}) usamos a linearidade
do tra\c{c}o parcial. J\'a a Eq. (\ref{eq:pt5}) \'e obtida a partir
de (\ref{eq:pt4}) usando as express\~oes para os coeficientes complexos
definidos acima em termos do produto interno entre os vetores base.
Rearranjando os termos da Eq. (\ref{eq:pt6}) ``fazemos aparecer''
os operadores unit\'arios de (\ref{eq:pt7}), o que leva \`a Eq.
(\ref{eq:pt8}). 

Prosseguindo definimos os elementos de matriz dos \emph{operadores
de Kraus} na base $\{|S_{k}\rangle\}$ em termos dos elementos de
matriz de $\hat{U}$ na base produto\footnote{Queremos chamar a aten\c{c}\~ao para o fato de n\~ao ser raro encontrarmos
os operadores de Kraus escritos da seguinte forma (veja por exemplo
a Ref. \cite{Nielsen=000026Chuang}): $\hat{K}_{l}=\langle E_{l}|\hat{U}|E_{0}\rangle.$
Embora o significado dessa equa\c{c}\~ao seja claro para quem est\'a
familiarizado com o assunto, evitamos utiliz\'a-la neste artigo pois,
al\'em do produto matricial $\hat{U}|E_{0}\rangle$ n\~ao ser bem
definido matematicamente, esse tipo de nota\c{c}\~ao pode confundir
sem necessidade aqueles que estudam o t\'opico pela primeira vez.}:
\begin{equation}
\langle S_{k}|\hat{K}_{l}|S_{m}\rangle:=\langle S_{k}E_{l}|\hat{U}|S_{m}E_{0}\rangle.\label{eq:KrausO}
\end{equation}
Essa rela\c{c}\~ao implica em $\langle S_{m}|\hat{K}_{l}^{\dagger}|S_{k}\rangle=\langle S_{m}E_{0}|\hat{U}^{\dagger}|S_{k}E_{l}\rangle$
e portanto
\begin{eqnarray}
\hat{\rho}_{t}^{S} & = & \sum_{klmpr}\langle S_{k}|\hat{K}_{l}|S_{m}\rangle|S_{k}\rangle\langle S_{m}|\hat{\rho}^{S}|S_{r}\rangle\langle S_{p}|\langle S_{r}|\hat{K}_{l}^{\dagger}|S_{p}\rangle\nonumber \\
 & = & \sum_{klmpr}|S_{k}\rangle\langle S_{k}|\hat{K}_{l}|S_{m}\rangle\langle S_{m}|\hat{\rho}^{S}|S_{r}\rangle\langle S_{r}|\hat{K}_{l}^{\dagger}|S_{p}\rangle\langle S_{p}|\nonumber \\
 & = & \sum_{l}\mathbb{I}_{S}\hat{K}_{l}\mathbb{I}_{S}\hat{\rho}^{S}\mathbb{I}_{S}\hat{K}_{l}^{\dagger}\mathbb{I}_{S}.
\end{eqnarray}
Chegamos assim \`a \textit{representa\c{c}\~ao da soma de operadores
de Kraus}:
\begin{equation}
\hat{\rho}_{t}^{S}=\sum_{l}\hat{K}_{l}\hat{\rho}^{S}\hat{K}_{l}^{\dagger}.\label{eq:rhot1}
\end{equation}

\'E bastante comum encontrarmos o operador densidade $\hat{\rho}_{t}^{S}$
escrito como $\$[\hat{\rho}^{S}]$ (ou com outro s\'imbolo no lugar
de $\$$). Esse procedimento enfatiza que a opera\c{c}\~ao qu\^antica
$\$$ \'e um \emph{mapa} (discreto) entre os estados $\hat{\rho}^{S}$
e $\hat{\rho}_{t}^{S}$, o que \'e denotado por
\begin{equation}
\$:\hat{\rho}^{S}\mapsto\hat{\rho}_{t}^{S}.
\end{equation}
Como $\$$ \'e um mapa que leva operador em operador ele \'e, algumas
vezes, chamado de \emph{super-operador}. No \hyperref[propriedades]{Ap\^endice}
mostramos que os operadores de Kraus da Eq. (\ref{eq:KrausO}) est\~ao
definidos em $\mathcal{H}_{S}$, que s\~ao lineares e que a din\^amica
gerada por eles (Eq. (\ref{eq:rhot1})) preserva o tra\c{c}o e \'e
positiva. Tamb\'em discutimos a n\~ao  unicidade do conjunto de
operadores de Kraus que geram a evolu\c{c}\~ao temporal de $S$ e
o consequente limite no n\'umero de tais operadores necess\'arios
para descrev\^e-la.

\section{\'Atomo de dois n\'iveis interagindo com o v\'acuo do campo eletromagn\'etico}

\label{sec_atom}

Como o prop\'osito de exemplificar a aplica\c{c}\~ao da representa\c{c}\~ao
de Kraus, nesta se\c{c}\~ao vamos considerar um \'atomo de dois
n\'iveis, e.g. um \'atomo de Rydberg \cite{Gallagher,Gallas,Nussenzveig,Noorden}
ou sistemas an\'alogos \cite{Kastner,Recher,Nori_AAinS}, cujos estados
fundamental e excitado ser\~ao denotados, respectivamente, por $|S_{0}\rangle$
e $|S_{1}\rangle$. O ambiente desse \'atomo \'e o campo eletromagn\'etico,
que est\'a inicialmente no estado de v\'acuo (i.e., com nenhuma
excita\c{c}\~ao), que chamaremos de $|E_{0}\rangle$. Estados do
ambiente contendo $j$ f\'otons espraiados por seus modos s\~ao
denotados por $|E_{j}\rangle$. Sob essas condi\c{c}\~oes um processo
de emiss\~ao ``espont\^anea'' ocorrer\'a por causa da intera\c{c}\~ao
do sistema at\^omico com as flutua\c{c}\~oes qu\^anticas do v\'acuo
\cite{Agarwal,Glauber}. Essa din\^amica pode ser modelada via o
seguinte mapa fenomenol\'ogico \cite{Nielsen=000026Chuang,Preskill}:
\begin{eqnarray}
 &  & \hat{U}_{CAA}|S_{0}E_{0}\rangle=|S_{0}E_{0}\rangle,\label{eq:CAA}\\
 &  & \hat{U}_{CAA}|S_{1}E_{0}\rangle=\sqrt{1-p}|S_{1}E_{0}\rangle+\sqrt{p}|S_{0}E_{1}\rangle,\nonumber 
\end{eqnarray}
que em CIQ \'e conhecido como \emph{canal de amortecimento de amplitude}
(CAA) \cite{Nielsen=000026Chuang,Preskill}. A interpreta\c{c}\~ao
desse mapa \'e simples. Se o \'atomo est\'a inicialmente no estado
fundamental, ele continuar\'a assim pois n\~ao h\'a energia alguma
para ``circular'' pelos sistemas. No entando, se o \'atomo estiver
inicialmente no estado excitado, este emitir\'a, em um processo rand\^omico,
um f\'oton com \emph{probabilidade} $p$ (i.e., $0\le p\le1$), que
\'e proporcional ao tempo de intera\c{c}\~ao \'atomo-campo. Como
o ambiente \'e muito ``grande'', vai demorar tamb\'em um tempo muito
grande, considerado na pr\'atica como sendo infinito, para que o
f\'oton seja recapturado pelo \'atomo. Dessa observa\c{c}\~ao inferimos
que o estado assimpt\'otico do \'atomo ser\'a $|S_{0}\rangle$,
independentemente de qual seja o seu estado inicial. A probabilidade
$p$ \'e muitas vezes chamada de tempo parametrizado \cite{Celeri_RedField,Maziero12,Walborn12,MazieroRRMN}.
Se observarmos a taxa $\gamma$ com que um n\'umero muito grande
desses \'atomos, todos independentes um do outro, perdem suas excita\c{c}\~oes,
temos que $p\approx1-\mathrm{e}^{-\gamma t}$ \cite{MazieroRRMN}.
Portando $t=0\mbox{ s}$ corresponde a $p=0$, enquanto $p=1$ seria
o limite assimpt\'otico $t\rightarrow\infty$.

\subsection{Operadores de Kraus para o CAA}

Nesta sub-se\c{c}\~ao utilizaremos a rela\c{c}\~ao  $\langle S_{k}|\hat{K}_{l}|S_{m}\rangle:=\langle S_{k}E_{l}|\hat{U}_{CAA}|S_{m}E_{0}\rangle$,
com a a\c{c}\~ao do operador unit\'ario dada pelo canal de amortecimento
de amplitude (Eqs. (\ref{eq:CAA})), para obter os operadores de Kraus
que geram essa din\^amica qu\^antica:
\begin{itemize}
\item $\hat{K}_{0}$:
\begin{eqnarray}
\langle S_{0}|\hat{K}_{0}|S_{0}\rangle & = & \langle S_{0}E_{0}|\hat{U}_{CAA}|S_{0}E_{0}\rangle\nonumber \\
 & = & \langle S_{0}E_{0}|S_{0}E_{0}\rangle=1.
\end{eqnarray}
\begin{eqnarray}
\langle S_{0}|\hat{K}_{0}|S_{1}\rangle & = & \langle S_{0}E_{0}|\hat{U}_{CAA}|S_{1}E_{0}\rangle\nonumber \\
 & = & \langle S_{0}E_{0}|(\sqrt{1-p}|S_{1}E_{0}\rangle+\sqrt{p}|S_{0}E_{1}\rangle)\nonumber \\
 & = & 0.
\end{eqnarray}
\begin{eqnarray}
\langle S_{1}|\hat{K}_{0}|S_{0}\rangle & = & \langle S_{1}E_{0}|\hat{U}_{CAA}|S_{0}E_{0}\rangle\nonumber \\
 & = & \langle S_{1}E_{0}|S_{0}E_{0}\rangle=0.
\end{eqnarray}
\begin{eqnarray}
\langle S_{1}|\hat{K}_{0}|S_{1}\rangle & = & \langle S_{1}E_{0}|\hat{U}_{CAA}|S_{1}E_{0}\rangle\nonumber \\
 & = & \langle S_{1}E_{0}|(\sqrt{1-p}|S_{1}E_{0}\rangle+\sqrt{p}|S_{0}E_{1}\rangle)\nonumber \\
 & = & \sqrt{1-p}.
\end{eqnarray}
i.e., na base $\{|S_{j}\rangle\}$
\begin{equation}
\hat{K}_{0}=\begin{bmatrix}1 & 0\\
0 & \sqrt{1-p}
\end{bmatrix}.\label{eq:K0}
\end{equation}

\item $\hat{K}_{1}$:
\begin{eqnarray}
\langle S_{0}|\hat{K}_{1}|S_{0}\rangle & = & \langle S_{0}E_{1}|\hat{U}_{CAA}|S_{0}E_{0}\rangle\nonumber \\
 & = & \langle S_{0}E_{1}|S_{0}E_{0}\rangle=0.
\end{eqnarray}
\begin{eqnarray}
\langle S_{0}|\hat{K}_{1}|S_{1}\rangle & = & \langle S_{0}E_{1}|\hat{U}_{CAA}|S_{1}E_{0}\rangle\nonumber \\
 & = & \langle S_{0}E_{1}|(\sqrt{1-p}|S_{1}E_{0}\rangle+\sqrt{p}|S_{0}E_{1}\rangle)\nonumber \\
 & = & \sqrt{p}.
\end{eqnarray}
\begin{eqnarray}
\langle S_{1}|\hat{K}_{1}|S_{0}\rangle & = & \langle S_{1}E_{1}|\hat{U}_{CAA}|S_{0}E_{0}\rangle\nonumber \\
 & = & \langle S_{1}E_{1}|S_{0}E_{0}\rangle=0.
\end{eqnarray}
\begin{eqnarray}
\langle S_{1}|\hat{K}_{1}|S_{1}\rangle & = & \langle S_{1}E_{1}|\hat{U}_{CAA}|S_{1}E_{0}\rangle\nonumber \\
 & = & \langle S_{1}E_{1}|(\sqrt{1-p}|S_{1}E_{0}\rangle+\sqrt{p}|S_{0}E_{1}\rangle)\nonumber \\
 & = & 0.
\end{eqnarray}
i.e., na base $\{|S_{j}\rangle\}$
\begin{equation}
\hat{K}_{1}=\begin{bmatrix}0 & \sqrt{p}\\
0 & 0
\end{bmatrix}.\label{eq:K1}
\end{equation}

\item $\hat{K}_{l\ge2}=\hat{0}$, com $\hat{0}$ sendo o operador nulo,
pois h\'a no m\'aximo um f\'oton no sistema como um todo e por
conseguinte $\langle S_{m}E_{l\ge2}|\hat{U}_{CAA}|S_{n}E_{0}\rangle=0.$
\end{itemize}

\subsection{Din\^amica da coer\^encia qu\^antica sob o CAA}

Segundo Feynman, a coer\^encia qu\^antica (CQ) \'e o aspecto mais
fundamental da mec\^anica qu\^antica \cite{Feynman}. A palavra coer\^encia
(ou mesmo CQ) se fez presente por v\'arios anos em \'optica qu\^antica
e em \'areas correlatas (veja e.g. \cite{Mandel-Wolf,Li} e suas
refer\^encias). Nos \'ultimos dois anos tem-se reconsiderado a CQ
do ponto de vista de teoria de recursos \cite{Devetak,Gour,Horodecki_E,Brandao_T,Veitch,Marvian,Brandao,Horodecki},
o que gerou um intensa e produtiva atividade de pesquisa sobre esse
tema \cite{Aberg,Baumgratz,Levi,Winter,Streltsov,Lostaglio,Xi,Pires}. 

Aqui, evitando demasiados detalhes, diremos que um sistema f\'isico
preparado em um certo estado $\hat{\rho}$ possui CQ em rela\c{c}\~ao
a uma dada base $\{|o_{j}\rangle\}$ (ou observ\'avel $\hat{O}=\sum_{j}o_{j}|o_{j}\rangle\langle o_{j}|$,
com $o_{j}\in\mathbb{R}$) se, quando representado naquela base, possuir
elementos de matriz fora da diagonal principal ($\langle o_{j}|\hat{\rho}|o_{k}\rangle$
com $j\ne k$) n\~ao nulos. Note que isso implica na exist\^encia
de incerteza qu\^antica em rela\c{c}\~ao a qual resultado ser\'a
obtido em uma medida de $\hat{O}$ \cite{MazieroRBEF}. A soma dos
m\'odulos desses elementos de matriz nos fornece um boa medida de
CQ \cite{Baumgratz}:
\begin{equation}
C(\hat{\rho})=\sum_{j\ne k}|\langle o_{j}|\hat{\rho}|o_{k}\rangle|.
\end{equation}

Vamos estudar como a din\^amica do sistema considerado nesta se\c{c}\~ao
influencia a sua CQ. Assumiremos que o \'atomo est\'a inicialmente
em um estado qualquer: 

\begin{equation}
\hat{\rho}^{S}=\frac{1}{2}\left(\mathbb{I}_{S}+\vec{r}\cdot\vec{\hat{\sigma}}\right),\label{eq:rhoi}
\end{equation}
onde $\vec{r}=(r_{1},r_{2},r_{3})$ \'e o vetor de Bloch, com $r_{j}=\mathrm{Tr}\left(\hat{\rho}^{S}\hat{\sigma}_{j}\right)\in\mathbb{R}$
sendo as ``polariza\c{c}\~oes'' para este estado, e $\vec{\hat{\sigma}}=(\hat{\sigma}_{1},\hat{\sigma}_{2},\hat{\sigma}_{3})$
com as matrizes de Pauli, na base $\{|S_{j}\rangle\}$, sendo escritas
como:
\begin{equation}
\hat{\sigma}_{1}=\begin{bmatrix}0 & 1\\
1 & 0
\end{bmatrix}\mbox{, }\hat{\sigma}_{2}=\begin{bmatrix}0 & -i\\
i & 0
\end{bmatrix}\mbox{, }\hat{\sigma}_{3}=\begin{bmatrix}1 & 0\\
0 & -1
\end{bmatrix}.
\end{equation}
Assim, na base $\{|S_{j}\rangle\}$,
\begin{eqnarray}
\hat{\rho}^{S} & = & \frac{1}{2}\begin{bmatrix}1+r_{3} & r_{1}-ir_{2}\\
r_{1}+ir_{2} & 1-r_{3}
\end{bmatrix}
\end{eqnarray}
e a CQ do estado inicial, em rela\c{c}\~ao a essa base, \'e dada
por: 
\begin{eqnarray}
C(\hat{\rho}^{S}) & = & 2^{-1}(|r_{1}-ir_{2}|+|r_{1}+ir_{2}|)\label{eq:CQ0}\\
 & = & \sqrt{r_{1}^{2}+r_{2}^{2}}.\label{eq:CQ1}
\end{eqnarray}
em que usamos $r_{1},r_{2}\in\mathbb{R}$ e, para $z\in\mathbb{C}$,
\begin{equation}
|z|=|z^{*}|=\sqrt{\mathrm{Re}(z)^{2}+\mathrm{Im}(z)^{2}}.
\end{equation}

Se aplicarmos as Eqs. (\ref{eq:K0}) e (\ref{eq:K1}) ao estado inicial
geral da Eq. (\ref{eq:rhoi}), o estado evolu\'ido do \'atomo,
\begin{equation}
\hat{\rho}_{p}^{S}=\sum_{j=0}^{1}\hat{K}_{j}\hat{\rho}^{S}\hat{K}_{j}^{\dagger},
\end{equation}
ser\'a
\begin{eqnarray}
\hat{\rho}_{p}^{S} & = & \frac{1}{2}\left(\sum_{j=0}^{1}\hat{K}_{j}\mathbb{I}_{S}\hat{K}_{j}^{\dagger}+\sum_{j=0}^{1}\hat{K}_{j}\vec{r}\cdot\vec{\hat{\sigma}}\hat{K}_{j}^{\dagger}\right)\\
 & = & \frac{1}{2}\left(\sum_{j=0}^{1}\hat{K}_{j}\hat{K}_{j}^{\dagger}+\sum_{k=1}^{3}r_{k}\sum_{j=0}^{1}\hat{K}_{j}\hat{\sigma}_{k}\hat{K}_{j}^{\dagger}\right).
\end{eqnarray}
Podemos verificar que
\begin{eqnarray}
 &  & \sum_{j=0}^{1}\hat{K}_{j}\hat{K}_{j}^{\dagger}=K_{0}K_{0}^{\dagger}+K_{1}K_{1}^{\dagger}\nonumber \\
 &  & =\begin{bmatrix}1 & 0\\
0 & \sqrt{1-p}
\end{bmatrix}\begin{bmatrix}1 & 0\\
0 & \sqrt{1-p}
\end{bmatrix}+\begin{bmatrix}0 & \sqrt{p}\\
0 & 0
\end{bmatrix}\begin{bmatrix}0 & 0\\
\sqrt{p} & 0
\end{bmatrix}\nonumber \\
 &  & =\begin{bmatrix}1 & 0\\
0 & (1-p)
\end{bmatrix}+\begin{bmatrix}p & 0\\
0 & 0
\end{bmatrix}=\begin{bmatrix}1 & 0\\
0 & 1
\end{bmatrix}+p\begin{bmatrix}1 & 0\\
0 & -1
\end{bmatrix}\nonumber \\
 &  & =\mathbb{I}_{S}+p\hat{\sigma}_{3}
\end{eqnarray}
e
\begin{eqnarray}
 &  & \sum_{j=0}^{1}\hat{K}_{j}\hat{\sigma}_{1}\hat{K}_{j}^{\dagger}=\hat{K}_{0}\hat{\sigma}_{1}\hat{K}_{0}^{\dagger}+\hat{K}_{1}\hat{\sigma}_{1}\hat{K}_{1}^{\dagger}\nonumber \\
 &  & =\begin{bmatrix}1 & 0\\
0 & \sqrt{1-p}
\end{bmatrix}\begin{bmatrix}0 & 1\\
1 & 0
\end{bmatrix}\begin{bmatrix}1 & 0\\
0 & \sqrt{1-p}
\end{bmatrix}\nonumber \\
 &  & +\begin{bmatrix}0 & \sqrt{p}\\
0 & 0
\end{bmatrix}\begin{bmatrix}0 & 1\\
1 & 0
\end{bmatrix}\begin{bmatrix}0 & 0\\
\sqrt{p} & 0
\end{bmatrix}\nonumber \\
 &  & =\begin{bmatrix}0 & 1\\
\sqrt{1-p} & 0
\end{bmatrix}\begin{bmatrix}1 & 0\\
0 & \sqrt{1-p}
\end{bmatrix}+\begin{bmatrix}\sqrt{p} & 0\\
0 & 0
\end{bmatrix}\begin{bmatrix}0 & 0\\
\sqrt{p} & 0
\end{bmatrix}\nonumber \\
 &  & =\begin{bmatrix}0 & \sqrt{1-p}\\
\sqrt{1-p} & 0
\end{bmatrix}+\begin{bmatrix}0 & 0\\
0 & 0
\end{bmatrix}\nonumber \\
 &  & =\sqrt{1-p}\begin{bmatrix}0 & 1\\
1 & 0
\end{bmatrix}\nonumber \\
 &  & =\hat{\sigma}_{1}\sqrt{1-p}.
\end{eqnarray}
Da mesma maneira podemos obter a rela\c{c}\~ao geral (para $k=1,2,3$):
\begin{equation}
\sum_{j=0}^{1}\hat{K}_{j}\hat{\sigma}_{k}\hat{K}_{j}^{\dagger}=\left(\sqrt{1-p}\right)^{(1+\delta_{3k})}\hat{\sigma}_{k}.
\end{equation}
Assim o estado evolu\'ido do \'atomo toma a seguinte forma: 
\begin{eqnarray}
\hat{\rho}_{p}^{S} & = & 2^{-1}(\mathbb{I}_{S}+p\hat{\sigma}_{3}+r_{1}\sqrt{1-p}\hat{\sigma}_{1}+r_{2}\sqrt{1-p}\hat{\sigma}_{2}\nonumber \\
 &  & \hspace{1em}\hspace{1em}+r_{3}(1-p)\hat{\sigma}_{3})\\
 & = & \frac{1}{2}\left(\mathbb{I}_{S}+\vec{r}_{p}\cdot\vec{\hat{\sigma}}\right),
\end{eqnarray}
com o vetor de Bloch evolu\'ido no tempo sendo 
\begin{eqnarray}
\vec{r}_{p} & = & \left(r_{1}(p),r_{2}(p),r_{3}(p)\right)\nonumber \\
 & = & \left(r_{1}\sqrt{1-p},r_{2}\sqrt{1-p},p+r_{3}(1-p)\right).
\end{eqnarray}

\begin{widetext}

\begin{figure}[H]
\begin{centering}
\includegraphics[scale=0.41]{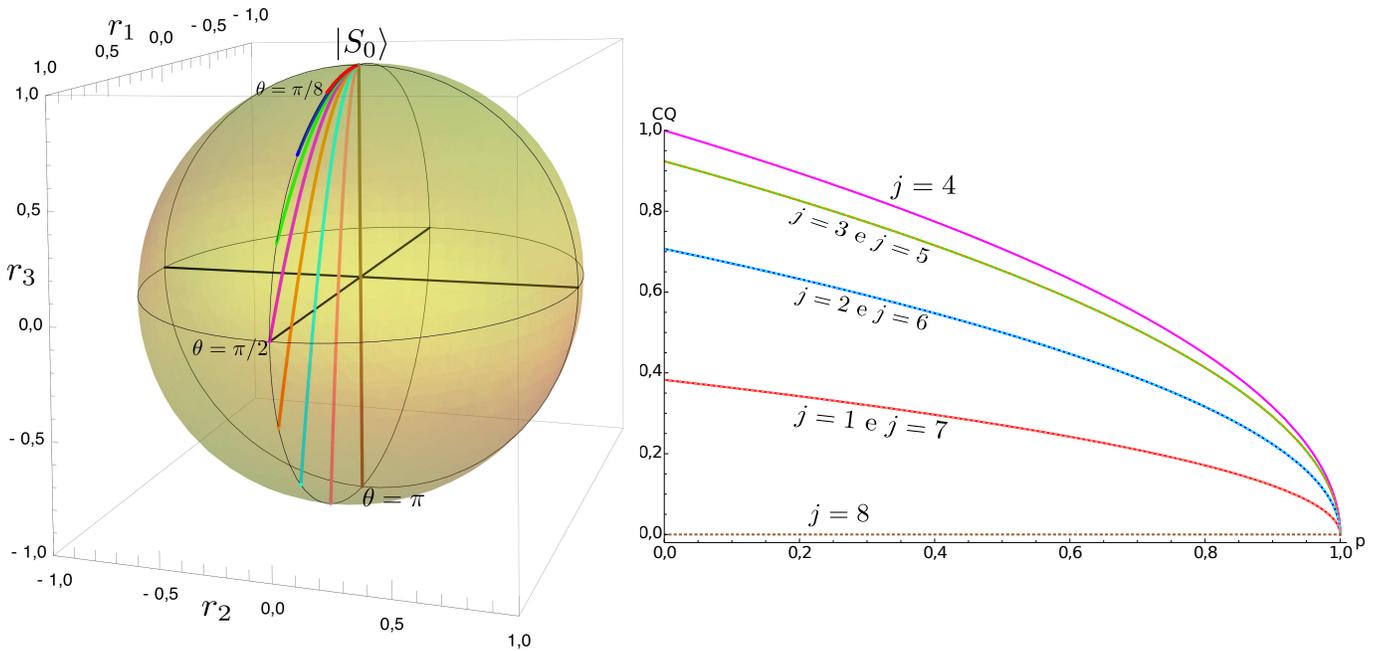}
\par\end{centering}

\caption{Qualquer estado $\hat{\rho}$ de um sistema de dois n\'iveis pode
ser representado por seu vetor de Bloch $\vec{r}$ em uma regi\~ao
de $\mathbb{R}^{3}$ conhecida como bola de Bloch. Em coordenadas
esf\'ericas, as componentes deste vetor s\~ao: $r_{1}=r\sin\theta\cos\phi$,
$r_{2}=r\sin\theta\sin\phi$ e $r_{3}=r\cos\theta$, com $r\in[0,1]$,
$\theta\in[0,\pi]$ e $\phi\in[0,2\pi)$. Nessa figura est\~ao mostradas
as evolu\c{c}\~oes temporais do vetor de Bloch (trajet\'orias no
gr\'afico de cima) e da coer\^encia qu\^antica (gr\'afico de baixo)
para estados iniciais puros ($r=1$, estes estados est\~ao na superf\'icie da bola de Bloch,
i.e., na esfera de Bloch) restritos ao plano $r_{2}=0$ ($\phi=0$),
com o \^angulo $\theta=j\pi/8$ ($j=1,\cdots,8$) e evolu\'idos
sob a a\c{c}\~ao do CAA. Estes estados iniciais s\~ao equivalentes
a $\hat{\rho}=|\psi\rangle\langle\psi|$ com $|\psi\rangle=\cos(\theta/2)|S_{0}\rangle+\sin(\theta/2)|S_{1}\rangle$.
No gr\'afico de cima obtemos os diferentes estados iniciais, e trajet\'orias,
iniciando com $\theta=\pi/8$ (curva vermelha) e aumentando $\theta$
de $\pi/8$ para obter o pr\'oximo estado mais abaixo. No gráfico
de baixo as linhas pontilhadas s\~ao para $j>4$, e para $j\le4$
usamos linhas cont\'inuas. Notamos que, para qualquer estado inicial,
o estado assimpt\'otico de $S$ \'e $|S_{0}\rangle$. Em todos os
casos a coer\^encia qu\^antica diminui monotonicamente com o tempo
parametrizado. No caso $j=8$ a CQ \'e zero sempre. Observamos tamb\'em
que a CQ \'e sim\'etrica em rela\c{c}\~ao ao \^angulo $\theta=\pi/2$,
ou seja, $C(\hat{\rho}_{p}^{S}(\pi/2+\delta))=C(\hat{\rho}_{p}^{S}(\pi/2-\delta))$
para qualquer \^angulo $\delta$ e tempo $p$. Isso indica a proporcionalidade
entre a CQ de um estado e a sua dist\^ancia euclidiana at\'e a linha
incoerente $q|S_{0}\rangle\langle S_{0}|+(1-q)|S_{1}\rangle\langle S_{1}|$,
com $0\le q\le1$.}

\label{fig_cq}
\end{figure}

\end{widetext}

Assim a CQ do \'atomo em um instante de tempo qualquer \'e dada
pela Eq. (\ref{eq:CQ1}) com $\{r_{j}\}_{j=1}^{2}$ substitu\'idos
por $\{r_{j}(p)\}_{j=1}^{2}$. Por conseguinte
\begin{equation}
C(\hat{\rho}_{p}^{S})=\sqrt{1-p}C(\hat{\rho}^{S}).
\end{equation}
Ou seja, se $C(\hat{\rho}^{S})\ne0$ a CQ do \'atomo diminui monotonicamente
indo a zero quando $t\rightarrow\infty$. A taxa de decaimento da
CQ n\~ao depende do estado inicial, mas sim da taxa de decaimento
do \'atomo; veja a Fig. \ref{fig_cq}.

\section{Considera\c{c}\~oes Finais}

Como foi bem colocado por Robert Sppekens em \cite{Spekkens}, existem
dois tipos de revolu\c{c}\~ao em f\'isica. A primeira, a mais notada,
\'e a aquela onde substitu\'imos uma teoria antiga por outra teoria
conceitualmente nova e mais abrangente, com a teoria da relatividade
geral de Einstein e a mec\^anica qu\^antica sendo importantes exemplos.
O segundo tipo de revolu\c{c}\~ao \'e mais lento, mas n\~ao menos
importante, e envolve uma mudan\c{c}a de perspectiva sobre uma teoria
existente. Um exemplo disso \'e o uso de princ\'ipios de simetria,
de m\'inima a\c{c}\~ao e termodin\^amicos em f\'isica. Tratando-se
teorias f\'isicas do ponto de vista de teoria de informa\c{c}\~ao,
ou seja, considerando-se \emph{estados f\'isicos }como sendo \emph{recursos},
tem-se conseguido aperfei\c{c}oar e generalizar esses princ\'ipios. 

Com o desenvolvimento da ci\^encia da informa\c{c}\~ao qu\^antica,
mostrou-se que os mais variados aspectos de estados f\'isicos (e.g.,
assimetria \cite{Marvian}, atermalidade \cite{Brandao_T}, emaranhamento
\cite{Horodecki_E}, n\~ao localidade qu\^antica \cite{Vicente},
coer\^encia qu\^antica \cite{Winter}, disc\'ordia qu\^antica \cite{Horodecki},
etc) podem ser utilizados como recursos para a realiza\c{c}\~ao mais
eficiente de diversos tipos de tarefas importantes para a ci\^encia,
tecnologia e sociedade. Nesse contexto, implementa\c{c}\~oes no laborat\'orio
de protocolos que se utilizem desses recursos envolvem um balan\c{c}o
delicado entre o controle necess\'ario do sistema qu\^antico de
interesse e da sua intera\c{c}\~ao com o ambiente. Por isso \'e imprescind\'ivel
levar em conta este \'ultimo quando da descri\c{c}\~ao do sistema
utilizado. Por conseguinte, \'e importante que, nas universidades
e institutos de educa\c{c}\~ao e pesquisa, trabalhemos antes e mais
com a teoria de sistemas qu\^anticos abertos.

Com a inten\c{c}\~ao de contribuir com esse projeto, neste artigo
derivamos, de forma simples e livre de ambiguidades, a representa\c{c}\~ao
de Kraus para a din\^amica de sistemas qu\^anticos abertos. Esperamos
que esse trabalho ajude a desmistificar esse formalismo e estimule
o seu uso, que se faz \'util para descrever muitos sistemas f\'isicos
reais de interesse para v\'arias \'areas de pesquisa e em particular
para CIQ. Nesse sentido, exemplificamos a aplica\c{c}\~ao desse formalismo
estudando a din\^amica do estado e da coer\^encia qu\^antica de
um \'atomo de dois n\'iveis interagindo com o v\'acuo do campo
eletromagn\'etico.

\selectlanguage{portuges}%
\global\long\def\acknowledgmentsname{Agradecimentos}

\selectlanguage{brazil}%
\begin{acknowledgments}
Este trabalho foi realizado com o suporte financeiro do CNPq, processos 441875/2014-9 e 303496/2014-2, do Instituto Nacional de Ci\^encia e Tecnologia de Informa\c{c}\~ao Qu\^antica, processo 2008/57856-6, e da CAPES, processo 6531/2014-08. Agrade\c{c}o a hospitalidade do Instituto de F\'isica e do Grupo de Espectroscopia Laser da Universidad de la Rep\'{u}blica, Uruguai, onde este artigo foi completado, e a Adriana Auyuanet por sugest\~oes e por discuss\~oes sobre o tema deste artigo. Agrade\c{c}o tamb\'em aos Revisores por suas sugest\~oes e cr\'iticas, que ajudaram a melhorar este artigo.
\end{acknowledgments}

\appendix*

\section{Propriedades da din\^amica gerada}

\label{propriedades}

Neste ap\^endice vamos verificar que os operadores de Kraus est\~ao
definidos em $\mathcal{H}_{S}$, que s\~ao lineares e que a din\^amica
gerada \'e positiva e preserva o tra\c{c}o. Al\'em disso, vamos
discutir a n\~ao  unicidade dos operadores de Kraus e o consequente
limite no n\'umero de tais operadores que s\~ao necess\'arios para
descrever a din\^amica de $S$.

\subsection{A\c{c}\~ao dos operadores de Kraus}

\label{sec_acao}

Vamos verificar que, como \'e esperado, os operadores de Kraus levam
vetores de $\mathcal{H}_{S}$ em vetores de $\mathcal{H}_{S}$, o
que se denota por $\hat{K_{l}}:\mathcal{H}_{S}\rightarrow\mathcal{H}_{S}$.
Seja $|\xi\rangle$ um vetor qualquer de $\mathcal{H}_{S}$. Usando
a resolu\c{c}\~ao da unidade escrevemos
\begin{eqnarray}
\hat{K}_{l}|\xi\rangle & = & \mathbb{I}_{S}\hat{K}_{l}\mathbb{I}_{S}|\xi\rangle\nonumber \\
 & = & \left(\sum_{j}|S_{j}\rangle\langle S_{j}|\right)\hat{K}_{l}\left(\sum_{k}|S_{k}\rangle\langle S_{k}|\right)|\xi\rangle\nonumber \\
 & = & \sum_{j,k}\langle S_{j}|\hat{K}_{l}|S_{k}\rangle\langle S_{k}|\xi\rangle|S_{j}\rangle\nonumber \\
 & = & \sum_{j,k}\langle S_{j}E_{l}|\hat{U}|S_{k}E_{0}\rangle\langle S_{k}|\xi\rangle|S_{j}\rangle\nonumber \\
 & = & \sum_{j}\langle S_{j}E_{l}|\hat{U}\left(\sum_{k}|S_{k}\rangle\langle S_{k}|\right)|\xi\rangle\otimes|E_{0}\rangle|S_{j}\rangle\nonumber \\
 & = & \sum_{j}\langle S_{j}E_{l}|\hat{U}|\xi E_{0}\rangle|S_{j}\rangle\label{eq:aKl}\\
 & := & \sum_{j}c_{j\xi}^{(l)}|S_{j}\rangle,\nonumber 
\end{eqnarray}
com os coeficientes complexos definidos como $c_{j\xi}^{(l)}:=\langle S_{j}E_{l}|\hat{U}|\xi E_{0}\rangle.$
Como qualquer combina\c{c}\~ao linear dos vetores de uma base de
$\mathcal{H}_{S}$ tamb\'em \'e um vetor de $\mathcal{H}_{S}$ vemos
que $\hat{K}_{l}|\xi\rangle\in\mathcal{H}_{S}$ e por conseguinte
de fato $\hat{K_{l}}:\mathcal{H}_{S}\rightarrow\mathcal{H}_{S}$ para
qualquer valor de $l$.

De forma similar, podemos escrever $\hat{K}_{l}^{\dagger}|\xi\rangle=\sum_{j}d_{j\xi}^{(l)}|S_{j}\rangle,$
com $d_{j\xi}^{(l)}=\langle S_{j}E_{0}|\hat{U}^{\dagger}|\xi E_{l}\rangle,$
para ver que $\hat{K}_{l}^{\dagger}|\xi\rangle\in\mathcal{H}_{S}\mbox{ }\forall l$.
Esse fato ser\'a \'util e.g. para a subsequente an\'alise da positividade
da din\^amica gerada.

\subsection{Linearidade dos operadores de Kraus}

Consideremos uma combina\c{c}\~ao linear de um conjunto qualquer
de vetores $|\zeta_{m}\rangle\in\mathcal{H}_{S}$ (n\~ao necessariamente
ortogonais), $|\xi\rangle:=\sum_{m}c_{m}|\zeta_{m}\rangle$ com $c_{m}\in\mathbb{C}$.
Um operador $\hat{L}:\mathcal{H}_{S}\rightarrow\mathcal{H}_{S}$ \'e
dito linear se $\hat{L}\left(\sum_{m}c_{m}|\zeta_{m}\rangle\right)=\sum_{m}c_{m}\hat{L}|\zeta_{m}\rangle$.
Para os operadores de Kraus, podemos utilizar a Eq. (\ref{eq:aKl}),
e a linearidade de $\hat{U}$ e do produto interno, para escrever
\begin{eqnarray}
\hat{K}_{l}\left(\sum_{m}c_{m}|\zeta_{m}\rangle\right) & = & \hat{K}_{l}|\xi\rangle\nonumber \\
 & = & \sum_{j}\langle S_{j}E_{l}|\hat{U}|\xi E_{0}\rangle|S_{j}\rangle\nonumber \\
 & = & \sum_{j}\langle S_{j}E_{l}|\hat{U}\sum_{m}c_{m}|\zeta_{m}\rangle\otimes|E_{0}\rangle|S_{j}\rangle\nonumber \\
 & = & \sum_{j}\langle S_{j}E_{l}|\hat{U}\sum_{m}c_{m}|\zeta_{m}E_{i}\rangle|S_{j}\rangle\nonumber \\
 & = & \sum_{m}c_{m}\sum_{j}\langle S_{j}E_{l}|\hat{U}|\zeta_{m}E_{i}\rangle|S_{j}\rangle\nonumber \\
 & = & \sum_{m}c_{m}\hat{K}_{l}|\zeta_{m}\rangle.
\end{eqnarray}

De forma similar pode-se mostrar que $\hat{K}_{l}^{\dagger}\left(\sum_{m}c_{m}|\zeta_{m}\rangle\right)=\sum_{m}c_{m}\hat{K}_{l}^{\dagger}|\zeta_{m}\rangle.$
Em suma, nestas duas primeiras sub-se\c{c}\~oes deste Ap\^endice
mostramos que os operadores de Kraus $\hat{K}_{l}$, e $\hat{K}_{l}^{\dagger}$,
s\~ao operadores lineares e que est\~ao definidos em $\mathcal{H}_{S}$.

\subsection{A din\^amica gerada preserva positividade}

Vamos mostrar que $\hat{\rho}_{t}^{S}$ \'e positivo semi-definido.
Para tal, comecemos considerando um vetor qualquer $|\xi\rangle\in\mathcal{H}_{S}$,
para o qual, pela linearidade do produto matricial,
\begin{equation}
\langle\xi|\hat{\rho}_{t}^{S}|\xi\rangle=\sum_{l}\langle\xi|\hat{K}_{l}\hat{\rho}^{S}\hat{K}_{l}^{\dagger}|\xi\rangle.
\end{equation}
Mostramos acima que $\hat{K}_{l}^{\dagger}:\mathcal{H}_{S}\rightarrow\mathcal{H}_{S}$.
Usemos essa propriedade aqui para definir $\hat{K}_{l}^{\dagger}|\xi\rangle=:|\xi_{l}\rangle\in\mathcal{H}_{S}$,
que implica em $\langle\xi_{l}|=|\xi_{l}\rangle^{\dagger}=\langle\xi|\hat{K}_{l}$,
e assim escrever 
\begin{eqnarray}
\langle\xi|\hat{\rho}_{t}^{S}|\xi\rangle & = & \sum_{l}\underset{\ge0}{\underbrace{\langle\xi_{l}|\hat{\rho}^{S}|\xi_{l}\rangle}}\label{eq:pos2}\\
 & \ge & 0.\nonumber 
\end{eqnarray}
Na Eq. (\ref{eq:pos2}) utilizamos a positividade do operador densidade
inicial de $S$.

\subsection{A din\^amica gerada preserva o tra\c{c}o}

Al\'em da positividade demonstrada anteriormente, para que uma distribui\c{c}\~ao
de probabilidades v\'alida seja gerada, o operador densidade deve
ter tra\c{c}o igual a um para todos os instantes de tempo $t$. Temos
que
\begin{eqnarray}
\mathrm{Tr}(\hat{\rho}_{t}^{S}) & = & \mathrm{Tr}\left({\textstyle \sum_{l}}\hat{K}_{l}\hat{\rho}^{S}\hat{K}_{l}^{\dagger}\right)\nonumber \\
 & = & {\textstyle \sum_{l}}\mathrm{Tr}\left(\hat{K}_{l}\hat{\rho}^{S}\hat{K}_{l}^{\dagger}\right)\nonumber \\
 & = & {\textstyle \sum_{l}}\mathrm{Tr}\left(\hat{K}_{l}^{\dagger}\hat{K}_{l}\hat{\rho}^{S}\right)\nonumber \\
 & = & \mathrm{Tr}\left(\left({\textstyle \sum_{l}}\hat{K}_{l}^{\dagger}\hat{K}_{l}\right)\hat{\rho}^{S}\right).
\end{eqnarray}
Na \'ultima equa\c{c}\~ao utilizamos a linearidade da fun\c{c}\~ao tra\c{c}o
e a sua propriedade c\'iclica. Agora usemos a resolu\c{c}\~ao da
unidade para ver que
\begin{eqnarray}
 &  & {\textstyle \sum_{l}}\hat{K}_{l}^{\dagger}\hat{K}_{l}={\textstyle \sum_{l}}\mathbb{I}_{S}\hat{K}_{l}^{\dagger}\mathbb{I}_{S}\hat{K}_{l}\mathbb{I}_{S}\nonumber \\
 &  & =\sum_{lmno}|S_{m}\rangle\langle S_{m}|\hat{K}_{l}^{\dagger}|S_{n}\rangle\langle S_{n}|\hat{K}_{l}|S_{o}\rangle\langle S_{o}|\nonumber \\
 &  & =\sum_{lmno}|S_{m}\rangle\langle S_{m}E_{0}|\hat{U}^{\dagger}|S_{n}E_{l}\rangle\langle S_{n}E_{l}|\hat{U}|S_{o}E_{0}\rangle\langle S_{o}|\nonumber \\
 &  & =\sum_{mo}|S_{m}\rangle\langle S_{m}E_{0}|\hat{U}^{\dagger}\sum_{nl}|S_{n}E_{l}\rangle\langle S_{n}E_{l}|\hat{U}|S_{o}E_{0}\rangle\langle S_{o}|\nonumber \\
 &  & =\sum_{mo}|S_{m}\rangle\langle S_{m}E_{0}|\hat{U}^{\dagger}\hat{U}|S_{o}E_{0}\rangle\langle S_{o}|\nonumber \\
 &  & =\sum_{mo}|S_{m}\rangle\langle S_{m}E_{0}|S_{o}E_{0}\rangle\langle S_{o}|\nonumber \\
 &  & =\sum_{mo}|S_{m}\rangle\delta_{mo}\langle S_{o}|\nonumber \\
 &  & ={\textstyle \sum}_{m}|S_{m}\rangle\langle S_{m}|\nonumber \\
 &  & =\mathbb{I}_{S}.
\end{eqnarray}
Esta igualdade garante\footnote{Vale observar que, em contraste como o formalismo geral de opera\c{c}\~oes qu\^anticas,
no contexto da din\^amica sistema-ambiente considerada aqui ${\textstyle \sum_{l}}\hat{K}_{l}^{\dagger}\hat{K}_{l}=\mathbb{I}_{S}$
\'e um fato e n\~ao um condi\c{c}\~ao a ser satisfeita pelos operadores
de Kraus para garantir que a din\^amica gerada preserve o tra\c{c}o.} que a din\^amica descrita acima preserva o tra\c{c}o do operador
densidade evolu\'ido do sistema, i.e., $\mathrm{Tr}(\hat{\rho}_{t}^{S})=\mathrm{Tr}\left(\hat{\rho}^{S}\right)=1,$
e portanto que gera um estado qu\^antico v\'alido para qualquer instante
de tempo.

\subsection{N\~ao unicidade dos operadores de Kraus}

O conjunto de operadores de Kraus que gera um mapa $\$:\hat{\rho}^{S}\rightarrow\hat{\rho}_{t}^{S}$
n\~ao \'e \'unico. Para verificar essa afirma\c{c}\~ao consideremos
\cite{Nielsen=000026Chuang}:
\begin{equation}
\hat{K}_{l}':=\sum_{n}V_{ln}\hat{K}_{n}\label{eq:KK}
\end{equation}
com $V_{ln}$ sendo os elementos de matriz de um operador unitário
qualquer ($\hat{V}^{\dagger}\hat{V}=\mathbb{I}$). O estado de $S$
obtido usando-se os operadores $\hat{K}_{l}'$ \'e:
\begin{eqnarray}
\sum_{l}\hat{K}_{l}'\hat{\rho}^{S}\hat{K}_{l}'^{\dagger} & = & \sum_{l}\left(\sum_{n}V_{ln}\hat{K}_{n}\right)\hat{\rho}^{S}\left(\sum_{m}V_{lm}^{*}\hat{K}_{m}^{\dagger}\right)\nonumber \\
 & = & \sum_{nm}\sum_{l}V_{lm}^{*}V_{ln}\hat{K}_{n}\hat{\rho}^{S}\hat{K}_{m}^{\dagger}\nonumber \\
 & = & \sum_{n}\hat{K}_{n}\hat{\rho}^{S}\hat{K}_{n}^{\dagger}\nonumber \\
 & = & \hat{\rho}_{t}^{S}
\end{eqnarray}
pois
\[
\sum_{l}V_{lm}^{*}V_{ln}=\sum_{l}(\hat{V}^{\dagger})_{ml}V_{ln}=(\hat{V}^{\dagger}\hat{V})_{mn}=\delta_{mn}.
\]
Ou seja, os conjuntos de operadores $\{\hat{K}_{l}'\}$ e $\{\hat{K}_{l}\}$
geram a mesma din\^amica qu\^antica para $S$.

No contexto da din\^amica sistema-ambiente que estamos interessados
aqui, uma rela\c{c}\~ao como a dada na Eq. (\ref{eq:KK}) \'e obtida
se, \emph{depois} de o sistema e ambiente evolu\'irem sob a a\c{c}\~ao
de $\hat{U}$, uma transforma\c{c}\~ao unit\'aria $\hat{V}$ for
aplicada ao ambiente. Assim, a composi\c{c}\~ao
\begin{equation}
\hat{U}'=(\mathbb{I}_{S}\otimes\hat{V})\hat{U}
\end{equation}
nos leva aos operadores de Kraus:
\begin{eqnarray}
 &  & \langle S_{k}|\hat{K}_{l}'|S_{m}\rangle=\langle S_{k}E_{l}|\hat{U}'|S_{m}E_{0}\rangle\nonumber \\
 &  & =\langle S_{k}E_{l}|(\mathbb{I}_{S}\otimes\hat{V})\hat{U}|S_{m}E_{0}\rangle\nonumber \\
 &  & =\langle S_{k}E_{l}|(\mathbb{I}_{S}\otimes\hat{V})(\mathbb{I}_{S}\otimes{\textstyle \sum_{n}}|E_{n}\rangle\langle E_{n}|)\hat{U}|S_{m}E_{0}\rangle\nonumber \\
 &  & =\sum_{n}(\langle S_{k}|\otimes\langle E_{l}|\hat{V}|E_{n}\rangle\langle E_{n}|)\hat{U}|S_{m}E_{0}\rangle\nonumber \\
 &  & =\sum_{n}V_{ln}\langle S_{k}E_{n}|\hat{U}|S_{m}E_{0}\rangle\nonumber \\
 &  & =\sum_{n}V_{ln}\langle S_{k}|\hat{K}_{n}|S_{m}\rangle,
\end{eqnarray}
que \'e equivalente \`a Eq. (\ref{eq:KK}). Vale observar que essas
rota\c{c}\~oes locais, aplicadas depois da din\^amica conjunta,
raramente t\^em alguma implica\c{c}\~ao relevante nas fun\c{c}\~oes
que consideramos em ci\^encia da informa\c{c}\~ao qu\^antica. Ou
seja, neste contexto podemos, em geral, desconsiderar a n\~ao unicidade
dos operadores de Kraus.

\subsection{Limite no n\'umero de operadores de Kraus necess\'arios para descrever
uma din\^amica qu\^antica}

Como a dimens\~ao do ambiente (e de $\hat{U}$) n\~ao \'e limitada,
n\~ao h\'a um motivo aparente para esperarmos que o n\'umero de
operadores de Kraus deva ser limitado. Não obstante, vamos mostrar
que um conjunto de at\'e $d_{S}^{2}$ operadores de Kraus \'e suficiente
para gerar qualquer din\^amica de um sistema com dimens\~ao $d_{S}$. 

Os operadores de Kraus est\~ao definidos em $\mathcal{H}_{S}$ (i.e.,
$\hat{K}_{l}:\mathcal{H}_{S}\rightarrow\mathcal{H}_{S}$) e, por conseguinte,
podem ser representados por matrizes complexas com dimens\~ao $d_{S}\mathrm{x}d_{S}$.
O espa\c{c}o formado por essas matrizes \'e denotado por $\mathbb{C}^{d_{S}\mathrm{x}d_{S}}$.
Seguindo a Ref. \cite{Nielsen=000026Chuang}, vamos come\c{c}ar mostrando
que para a matriz Hermitiana $\hat{W}$ definida por
\begin{equation}
W_{lm}:=\mathrm{Tr}(\hat{K}_{l}^{\dagger}\hat{K}_{m})
\end{equation}
temos que $\mathrm{rank}(\hat{W})\le d_{S}^{2}$, em que o rank de
$\hat{W}$ \'e definido como o n\'umero de seus vetores coluna que
s\~ao linearmente independentes (LI) \cite{Horn}. Para isso, notemos
que existe uma base ortonormal para $\mathbb{C}^{d_{S}\mathrm{x}d_{S}}$
com $d_{S}^{2}$ elementos, o que implica que n\~ao mais do que $d_{S}^{2}$
dos operadores $\hat{K}_{l}$ podem ser LI. Suponhamos, sem perda
de generalidade, que os primeiros $d_{S}^{2}$ operadores $\hat{K}_{l}$
s\~ao LI. Com isso podemos escrever, para $m>d_{S}^{2}$, a combina\c{c}\~ao
linear: 
\begin{equation}
\hat{K}_{m}={\textstyle \sum_{l=1}^{d_{S}^{2}}}c_{l}^{(m)}\hat{K}_{l},
\end{equation}
com $c_{l}^{(m)}\in\mathbb{C}$. Pode-se verificar facilmente que
isso implica que o $m$-\'esimo vetor coluna de $\hat{W}$, para
$m>d_{S}^{2}$, \'e uma combina\c{c}\~ao linear dos $d_{S}^{2}$
primeiros vetores coluna:
\begin{equation}
W_{lm}={\textstyle \sum_{j=1}^{d_{S}^{2}}}c_{j}^{(m)}W_{lj}.
\end{equation}
Fica provado assim que
\begin{equation}
\mathrm{rank}(\hat{W})\le d_{S}^{2}.
\end{equation}
Utilizando esse resultado e o fato de que $\hat{W}$ \'e uma matriz
\emph{Hermitiana}, podemos diagonaliz\'a-la via uma transforma\c{c}\~ao unit\'aria
$\hat{V}$:
\[
\hat{W}':=\hat{V}\hat{W}\hat{V}^{\dagger}.
\]
Os $d_{S}^{2}$ elementos diagonais possivelmente n\~ao nulos de
$\hat{W}'$ s\~ao obtidos como segue:
\begin{eqnarray}
W_{jj}' & = & (\hat{V}\hat{W}\hat{V}^{\dagger})_{jj}=\sum_{l}V_{jl}(\hat{W}\hat{V}^{\dagger})_{lj}\nonumber \\
 & = & \sum_{lm}V_{jl}W_{lm}(\hat{V}^{\dagger})_{mj}=\sum_{lm}V_{jl}\mathrm{Tr}(\hat{K}_{l}^{\dagger}\hat{K}_{m})V_{jm}^{*}\nonumber \\
 & = & \mathrm{Tr}\sum_{l}V_{jl}\hat{K}_{l}^{\dagger}\sum_{m}V_{jm}^{*}\hat{K}_{m}=\mathrm{Tr}(\hat{K}_{j}'^{\dagger}\hat{K}_{j}').
\end{eqnarray}
Temos assim um conjunto com at\'e $d_{S}^{2}$ operadores de Kraus
$\hat{K}_{j}'=\sum_{m}V_{jm}^{*}\hat{K}_{m}$, que geram a mesma din\^amica
que o conjunto $\hat{K}_{m}$.

\end{document}